\documentclass[aps,prl,reprint,superscriptaddress,amsmath,amssymb,floatfix]{revtex4-2}
\usepackage[T1]{fontenc}
\usepackage{graphicx}
\usepackage{microtype}
\usepackage[colorlinks=true,citecolor=blue,urlcolor=blue,linkcolor=blue]{hyperref}
\newcommand{\ket}[1]{\lvert #1\rangle}
\newcommand{\bra}[1]{\langle #1\rvert}

\begin{document}
\title{Bulk and edge states in Fibonacci-Bose-Hubbard clusters}
\author{Fedorov G.P.}
\affiliation{Moscow Center for Advanced Studies, Moscow 123592, Russia}
\affiliation{Kotel'nikov Institute of Radio Engineering and Electronics, Russian Academy of Sciences, Moscow 125009, Russia}
\author{Konopleva E.E.}
\affiliation{Center for Engineering Physics, Moscow 121205, Russia}
\affiliation{Moscow Center for Advanced Studies, Moscow 123592, Russia}
\author{Astafiev O.V.}
\affiliation{Center for Engineering Physics, Moscow 121205, Russia}
\affiliation{Moscow Center for Advanced Studies, Moscow 123592, Russia}
\date{\today}
\begin{abstract}
We study the one- and two-boson energy subspaces of finite open Bose--Hubbard chains with Fibonacci-modulated hopping and zero on-site potential. Their spectra exhibit a hierarchy of bands and gaps, and a critical zero-energy mode for odd lengths, in agreement with prior research. For the singe-excitation subspace we find two families of length-invariant finite-energy states localized at the left cluster boundary. For the two-boson subspace, we report both doublon and separated-pair excitations in the form of Tamm/Shockley states localized near either end. In particular, the right-edge doublon states occur across all Fibonacci generations whereas the left-side states appear every second generation. We check numerically that the second-order perturbation theory accounts well for the structure of the spectrum and the edge states in the doublon subspace in the strong-coupling regime. The results are useful for future explorations of many-body physics beyond numerical tractability, specifically using near-term Josephson quantum simulators.
\end{abstract}
\maketitle

The Bose-Hubbard (BH) model describes multipartite coupled oscillatory systems with nonlinearity~\cite{kolovsky2016,jaksch1998}. In most cases, it is not solvable analytically, and at large scales may thus be useful to cast light upon several important physical and philosophical problems like the eigenstate thermalization hypothesis (ETH)~\cite{kaufman2016}, quantum chaos~\cite{kolovsky2016}, and the emergence of classical reality. In recent years, this model has attracted significant attention since it can be physically implemented with very high accuracy using a fully quantum-mechanical medium of transmon-type superconducting artificial atoms~\cite{roushan2017,fedorov2021}. Moreover, the fabrication and control techniques now allow building transmons with up to twelve resolved levels; high-fidelity adjacent-level control and readout have been demonstrated in the lowest ten levels~\cite{wang2025}, thus opening exciting perspectives in terms of reaching classical intractability even with small-scale devices.

An important intermediate step in the development of quantum simulators is the benchmarking of structures with predictable behaviors, ensuring the ability to control how precise the simulation actually is. To date, transmon-based simulators with uniform or alternating coupling patterns have been realized and studied; experiments have shown good agreement with theoretical predictions~\cite{fedorov2021,besedin2021}. Another example of a great precision test is a nine-qubit array used to reproduce the features of the Hofstadter butterfly by a quasiperiodic on-site energy modulation \cite{roushan2017}. Based on these successes, it seems feasible to proceed towards experimental research employing the Josephson quantum simulators in completely unsolvable regimes. For instance, building a large-scale BH simulator with randomized couplings and running it at high occupation will allow physical exploration beyond both theory and numerics. We believe that, in the near term, this way towards classical intractability may become a viable alternative to the multi-orbital quantum simulators \cite{konopleva2026new,konopleva2026nontrivial}.

\begin{figure}[t]
	\centering
	\includegraphics[width=\columnwidth]{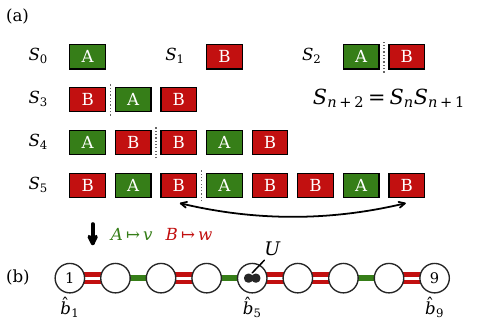}
	\caption{Construction of the Fibonacci-Bose-Hubbard chain. (a) Starting from $S_0=A$ and $S_1=B$, concatenate $S_{n+2}=S_nS_{n+1}$; dotted lines mark the joins. Green and red denote $A$ and $B$ bonds, respectively. The double-headed arrow indicates that the suffixes obtained by removing the first two letters are always inversion-symmetric (palindromes). For $n\geq2$, the first two letters are $AB$ or $BA$, while the right termination stays the same. (b) The letters set the hopping amplitudes $A\mapsto v$ and $B\mapsto w$ between bosonic modes $\hat b_j$. The illustrated word $S_5=BABABBAB$ has nine sites. The two dots at site 5 depict a same-site pair, which has interaction energy $U$ under Eq.~\eqref{eq:bh}.}
	\label{fig:construction}
\end{figure}

\begin{figure*}[tbp]
	\centering
	\includegraphics[width=\textwidth]{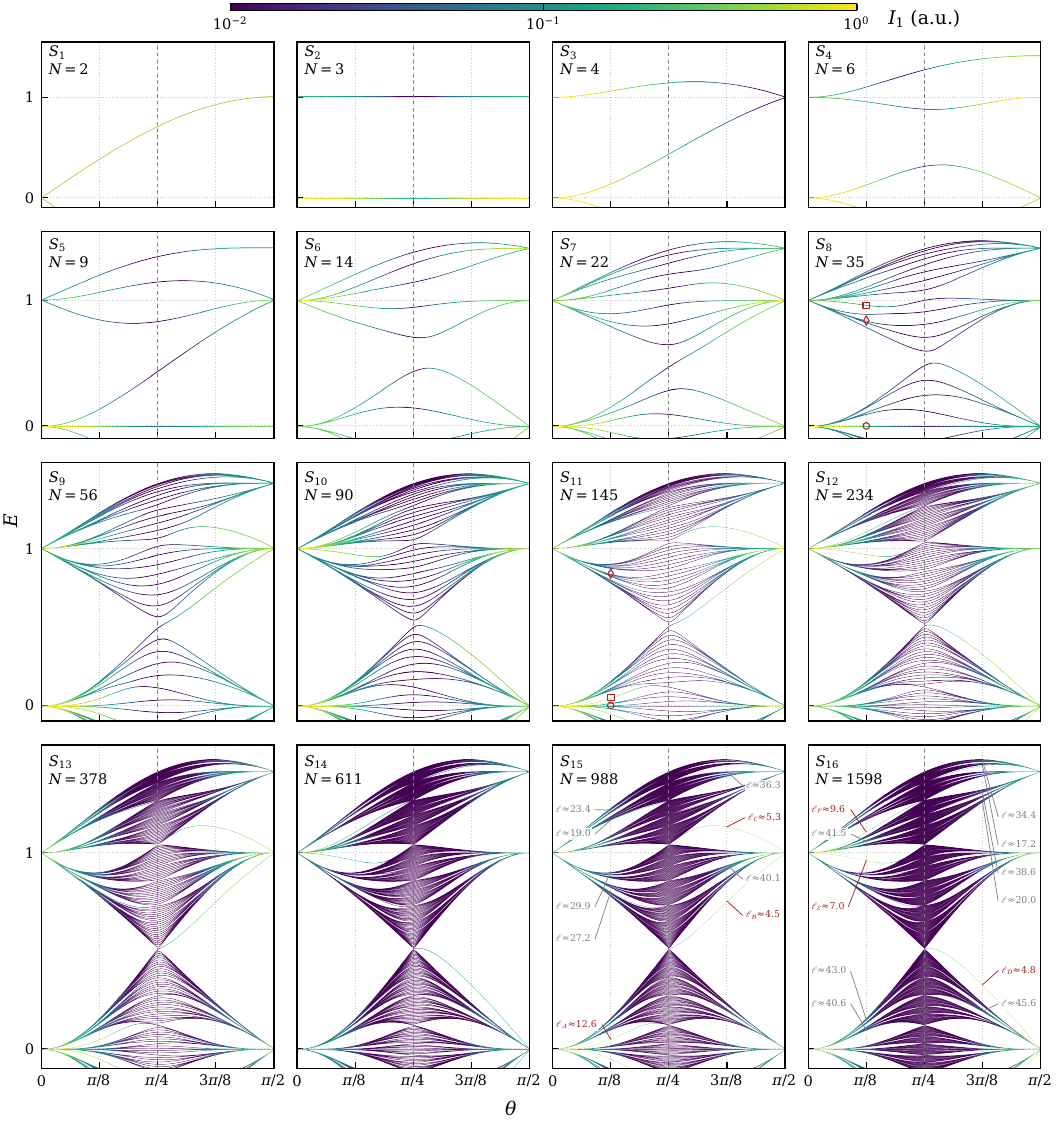}
	\caption{Nonnegative one-boson spectra of sixteen complete Fibonacci words, $S_1$ through $S_{16}$; the negative energies follow by reflection about $E=0$. The interaction term of Eq.~\eqref{eq:bh} vanishes in this sector. Line colors encode $I_1$ through the length-specific normalized coordinate defined in the text, clipped below $10^{-2}$. At $\theta=\pi/8$, $\{\circ,\Diamond,\square\}$ mark selected zero-energy, bulk, and localized states, respectively, for $S_{8,11}$. The gray dashed line at $\pi/4$ marks $v=w$ where the chain is uniform. The panels $S_{15,16}$ contain annotations for multiple exponentially left-edge localized states inside the bandgaps, including their localization lengths. The states in the largest gaps are named $\mathcal{A}$-$\mathcal{F}$.}
	\label{fig:spectra}
\end{figure*}
In this work, we show that resorting to complete disorder may not be necessary to reach interesting physics beyond classical computation; for example, one can study a number of pronounced physical effects in the domain of quasicrystals \cite{sandberg2024josephson} for which analytical solutions also do not exist. Beginning with a Su--Schrieffer--Heeger chain~\cite{ssh1979} and replacing some bonds by their complements, one can obtain the so-called Fibonacci chain,  a well-known one-dimensional quasicrystal. The models of this family are known to have hierarchical spectra and critical eigenstates \cite{kohmoto1983, kohmoto1987, jagannathan2021}. Since the SSH model zero-energy mode is topologically protected with respect to bond perturbations, one can expect an exact zero mode to remain in the Fibonacci chain.  Indeed, prior results show that it is observed, and its structure can be described analytically; however, this state is not edge-bound anymore \cite{mace2017}. Only Tamm-like boundary states in Fibonacci structures have been previously reported when an energy defect had been introduced at an edge of the palindrome subword \cite{zijlstra1999}. 

To check whether the characteristic quasicrystal phenomena are reachable in the moderately-sized clusters, we generate a sequence of complete Fibonacci words and see what the standard local Bose--Hubbard Hamiltonian leads to when the couplings are chosen accordingly. We analyze the one-boson sector first, then the two-boson spectra containing the bound (or quasi-independent) photon pairs and their edge states, which have been reported before in periodic dimerized chains \cite{gorlach2017, diliberto2016}. 

\begin{figure*}[tbp]
	\centering
	\includegraphics[width=\textwidth]{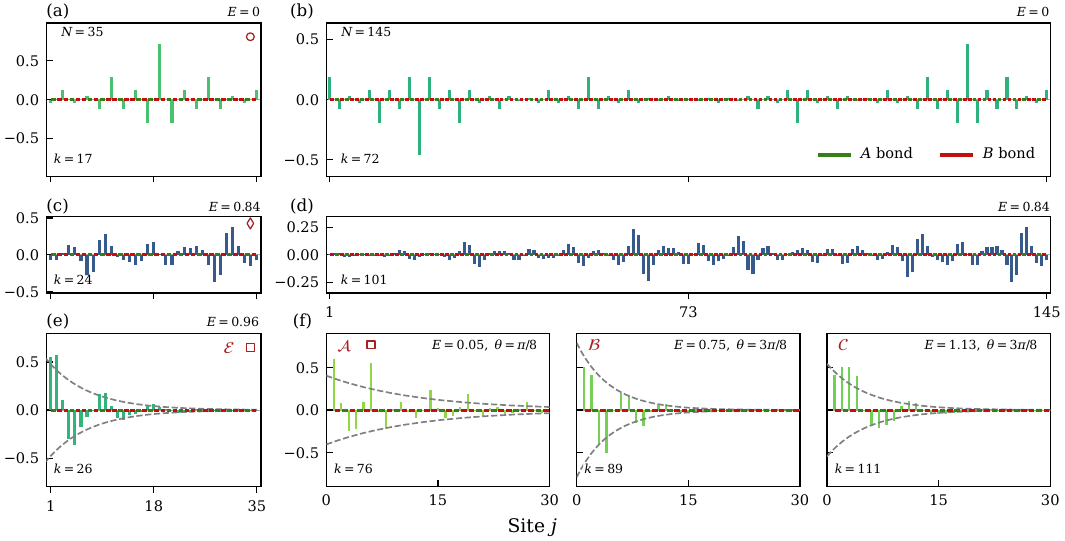}
	\caption{Visualization of selected one-boson states $\ket{\psi_k}$ for $N=35,145$ (left to right); stems show signed $\left<j|\psi_k\right>$ amplitudes. Panels (a--e) use $\theta=\pi/8$. The first row (a,b) contains the exact odd-chain zero modes, with vanishing amplitudes on even-numbered sites. The second row (c,d) shows spread-out profiles at $E\approx0.84$. Panel (e) shows the localized $\mathcal E$-type state. Under (f) are three $N=145$ profiles corresponding to $\mathcal A$, $\mathcal B$, and $\mathcal C$ in Fig.~\ref{fig:spectra}: $\theta=\pi/8$ for $\mathcal A$ and $3\pi/8$ for $\mathcal B,\mathcal C$. These profiles are cropped to $0\leq j\leq30$ retaining their full-chain normalization. Gray dashed curves in (e,f) show exponential fits to local maxima of $|\psi_j|$ as guides to the eye. Stem colors use the scale of Fig.~\ref{fig:spectra}.}
	\label{fig:states}
\end{figure*}

The complete words are generated by the rule $S_0=A,\quad S_1=B,\quad S_{n+2}=S_nS_{n+1}.$ For $n\geq2$, even and odd generations begin with $AB$ and $BA$, respectively; removing these two letters leaves an inversion-symmetric word \cite{zijlstra1999}. The letters label bonds, so a word of length $L_n$ defines $N=L_n+1$ bosonic sites, see Fig.~\ref{fig:construction}(a). Next, from any word, the corresponding Hamiltonian 
\begin{equation}
\hat H=\sum_{j=1}^{N-1}t_j(\hat b_j^\dagger\hat b_{j+1}+\mathrm{h.c.})+\frac{U}{2}\sum_{j=1}^{N}\hat n_j(\hat n_j-1)
	\label{eq:bh}
\end{equation}
follows, where the hopping amplitudes are fixed by $t_j=v$ for $A$ and $t_j=w$ for $B$. We use hereafter $v=\cos\theta$ and $w=\sin\theta$, $0\leq\theta\leq\pi/2$, obtaining a fixed hopping scale $v^2+w^2=1$. One can notice $\hat H(S_n, \theta) \equiv \hat H(\overline S_n, \pi/2-\theta)$, where $\overline S_{0,1,2,...}=\{B,\ A,\ BA,\ ...\}$ is the complementary word with exchanged strong and weak links. This means that our study covering the $0\leq\theta\leq\pi/2$ interval captures the Fibonacci clusters for both possible initial seeds.

\begin{figure*}[tbp]
	\centering
	\includegraphics[width=\textwidth]{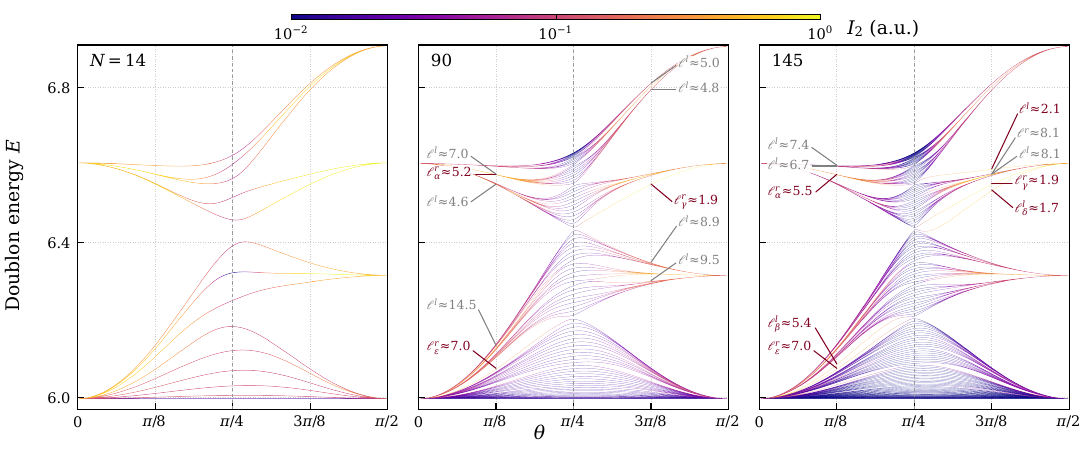}
	\caption{Repulsively bound two-boson (doublon) spectra of the complete Fibonacci words $S_{6,10,11}$ with $N=14, 90,145$, for $U=6$. The lower separated-pair bands are not shown (each panel shows the upper $N$ eigenstates of \eqref{eq:bh}); the same-site-pair probability remains above approximately $87.5\%$ over the plotted scan. Line color gives the full two-boson $I_2$ defined in the text using the same length-specific rescaling and logarithmic clipping described for Fig.~\ref{fig:spectra}. For $N=90,145$, annotations for $\theta=\pi/8$ and $3\pi/8$ identify several edge-states with the shortest $\ell^{l,r}$ which denote the amplitude decay length inward from the left or right edge. In the $N=145$ panel, subscripts $\alpha$--$\delta$ identify the four doublons shown in Fig.~\ref{fig:doublon_states}(i--l), respectively. Labels $\alpha,\gamma,\epsilon$ mark matching right-edge states in the adjacent $N=90,145$ generations; $\epsilon$ is the state just below $\beta$ at $\theta=\pi/8$.}
	\label{fig:doublon_spectra}
\end{figure*}

In the one-boson sector, the interaction term of Eq.~\eqref{eq:bh} vanishes. In the site basis $\ket{j}=\hat b_j^\dagger\ket{0}$, the Hamiltonian reduces to
\begin{equation}
 \hat H_1=\sum_{j=1}^{N-1}t_j\bigl(\ket{j}\bra{j+1}+\ket{j+1}\bra{j}\bigr).
 \label{eq:hamiltonian}
\end{equation}
Fig.~\ref{fig:spectra} shows numerically calculated spectra of $\hat H_1(S_{1-16},\theta)$ with colors indicating the measure of localization of each eigenstate $\ket{\psi_k}$. For the latter, we use a quantity known as the {inverse participation ratio}, $I_1 = \sum_{j=1}^{N}{|\left<j|\psi_k\right>|^4},\ I_1\in[N^{-1}, 1]$ ($I_1$ means that it is the quantity for the one-boson states). Since the lower extremum drops significantly at increased cluster lengths, here and below we rescale the colormaps for each panel to $[0,1]$, choosing the extremal values among all $\{\ket{\psi_k}, \theta\}$ for each $N$. This rescaled coordinate is displayed logarithmically with clipping below $10^{-2}$. 

One can note immediately that $\hat \Gamma \hat H_1\hat \Gamma=-\hat H_1$ for $\hat \Gamma\ket{j}=(-1)^{j-1}\ket{j}$, and thus the one-boson spectrum is chiral. At $\theta=0$, isolated $A$ bonds form dimers with energies $\pm1$, and the remaining sites have zero energy, hence the bands converge to three points. At $\theta=\pi/2$, $v=0$, while the surviving $B$-coupled dimers and trimers give energies $0,\pm1,\pm\sqrt2$. Between these limits, each generation resolves finer levels within the well-known spectral hierarchy~\cite{kohmoto1987,jagannathan2021}. At $\theta=\pi/4$, the chain is uniform, and the exact one-boson energies are $E_k=\sqrt2\cos[k\pi/(N+1)]$, $k=1,\ldots,N$.

Another analytical result concerns the zero-energy state. For even $N$, $\det \hat H_1 \neq 0$ can be explicitly calculated, excluding a zero eigenvalue away from the $\theta=\{0,\pi/2\}$ endpoints, which is clearly evident in Fig. \ref{fig:spectra}.  For odd $N$, a zero mode always exists, and, away from the decoupled endpoints, its profile at odd sites can be predicted by introducing $r=v/w$ and defining the bond indicator $\alpha_j=1 (0)$ for $A (B)$ bond:
\begin{equation}
\psi_{2m+1}=(-1)^m r^{h_m}\psi_1,\ 
 h_m=\sum_{k=1}^{m}(\alpha_{2k-1}-\alpha_{2k}).
 \label{eq:height}
\end{equation}
The height $h_m$ counts the excess of $A$ bonds in odd over even bond positions. Its variation along a Fibonacci word controls the zero-mode envelope~\cite{mace2017}; when $r=1$, every odd site has the same probability. The same expression is usually derived for an exponentially decaying zero-energy topologically protected edge state when the SSH model is considered.

\begin{figure}[tbp]
	\centering
	\includegraphics[width=\columnwidth]{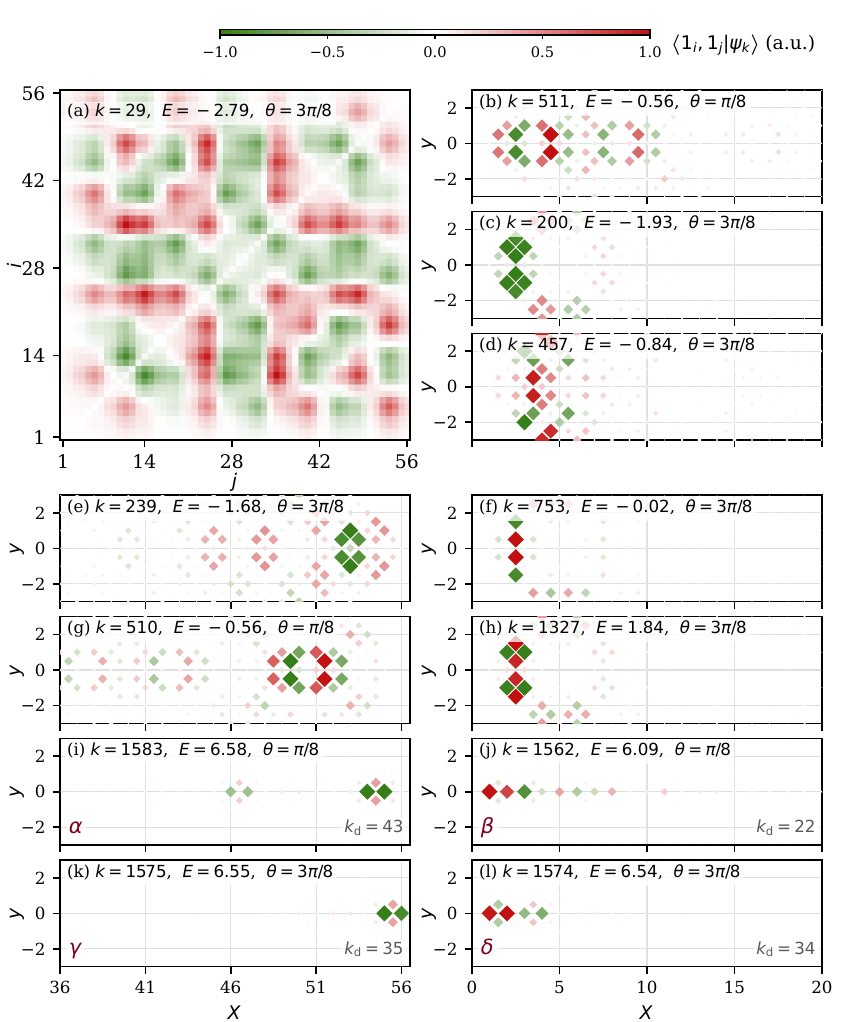}
\caption{Amplitudes of the two-boson states for $S_9$ ($N=56$), $U=6$, at the angles and energies indicated. States are defined by $k$, the zero-based index in the full 1596-state spectrum, and $k_{\rm d}$ indexing the doublon subband.  (a) A spread-out state in the lower band, shown as a symmetric matrix of normalized Fock coefficients (the largest amplitude is scaled to unity). (b--h) Edge-concentrated split-pair states, predominantly populating $i\ne j$ basis states; (i--l) doublon edge states. The rectangular panels use $X=(i+j)/2$ and $y=(i-j)/2$ to separate the diagonal zone of the configuration space. Right-edge profiles occupy (e,g,i,k); the other panels show left-edge profiles.}
	\label{fig:doublon_states}
\end{figure}

We numerically analyze this and two other types of eigenfunctions of \eqref{eq:hamiltonian} in Fig.~\ref{fig:states}. It presents signed one-boson amplitudes at $\theta=\{\pi/8,\ 3\pi/8\}$ for $S_{8,11}$ ($N=\{35,145\}$). The first row contains the exact zero modes,  demonstrating prominent self-similar interior peaks over the whole length of the longer words. These are critical, multifractal \cite{mace2017} states whose structure is neither uniformly extended nor exponentially localized. One can also notice that the amplitudes of the wavefunction only take discrete values, which is expected from \eqref{eq:height}. 

The row below depicts a family of delocalized bulk in-band states. Just as the zero mode, these states retain part of their structure from one generation to the next. This is ensured by the recurrence relation $t_{j-1}\psi_{j-1} = E_k\psi_j-t_{j} \psi_{j+1}$ as long as the energy $E_k$ is observed across generations. In spite of this fact, if $\ket{\psi_k}$ is a bulk state, all eigenstates of larger chains close in energy will become spread out more and more over the cluster.

The bottom row of Fig.~\ref{fig:states} shows several states with a relatively high localization. The selected lengths mark the first occurrences of stable Shockley-like edge-bound profiles at the left edge (no energy defect is required for these states, in contrast to the doublon case below). These states persist infinitely over later generations that have the same left edge, complementing previous reports ($\alpha=0$ case in Sec. VI of \cite{zijlstra1999}; note that right/left is exchanged there). The manifestation of a family of such wavefunctions in Fig. \ref{fig:spectra} is in the bright whiskers bending inside the bandgaps, copied over and over in odd and even generations. After fitting the logarithms of local maxima of $|\psi_j|$ with $f(j)=C - j/\ell$, we extract the localization lengths $\ell$ for some of these states and annotate them in Fig. \ref{fig:spectra}. The edge states in the largest bandgaps are labeled by $\ell_{\mathcal{A}-\mathcal{F}}$. We check numerically that all these states retain their shape in further generations with fidelity error lower than $10^{-10}$, and in the Supplementary we also provide the dependence of $\ell_{\mathcal{A}-\mathcal{F}}$ on $\theta$ (or $r$).  We also note that  stable localized states in the bulk obviously cannot exist for a Fibonacci chain, since after first emergence any such peak will get repeated exponentially many times in the following generations, consistent with theoretical results for infinite chains prohibiting the square-normalizable eigenfunctions \cite{damanik2008spectrum}. Finally, we report no edge states at the right edge, in accordance with \cite{zijlstra1999}.

Next, we turn to the two-boson subspace. It is known that the Bose-Hubbard model can describe effects of (i) binding the bosons together and (ii) their spatial confinement as a group. However, these effects do not necessarily occur at the same time. For example, in the uniform chain, same-site binding can occur for two (doublons), three (triplons) or more particles. Nevertheless, edge-localized states are not formed by doublons in the uniform case\cite{valiente2008}; edge localization occurs for clusters of three or more bosons \cite{pinto2009}. Dimerization, in a Peierls transition, can allow two-boson edge states; interestingly, they appear even for a topologically trivial configuration excluding the single-particle edge states \cite{gorlach2017, diliberto2016}. The SSH chain exhibits also some scattering states with one particle at an edge and the other extended~\cite{diliberto2016}. Adding nearest-neighbor longitudinal interactions produces some further edge states~\cite{diliberto2017}. We show that for the FBH model, one can observe edge localization for \textit{both} bound and separated pairs of bosonic excitations.

We diagonalize Eq.~\eqref{eq:bh} in the basis $\ket{j,j}=(\hat b_j^\dagger)^2\ket{0}/\sqrt2$ (doublons) and $\ket{i,j}=\hat b_i^\dagger\hat b_j^\dagger\ket{0}$ for $i<j$ (split pairs), of dimension $N(N+1)/2$. We choose repulsion $U=6$ for the numerical spectra as changing repulsion to attraction of equal magnitude in \eqref{eq:bh} reflects all energies about zero while preserving occupation probabilities, and then the bound-pair band lies below the scattering states, as relevant to attractive transmon implementations~\cite{fedorov2021,besedin2021}.

Fig.~\ref{fig:doublon_spectra} shows the upper $N$ eigenlevels of the two-boson eigenspace (the doublon band) for several Fibonacci generations. As before, one can observe the hierarchical splittings that become finer with increasing $N$.  The coloring is based on $I_2=\sum_{i\leq j}|\psi_{ij}|^4 \in [2/(N^2+N), 1],\ \psi_{ij}=\langle i,j|\psi_k\rangle$ for $i\leq j$, normalized as before. A second quantity $P_d=\sum_j|\psi_{jj}|^2$ measures the probability that both bosons occupy the same site. We observe for the doublon band $P_d\gtrsim 87.5\%$; $P_d$ is lower for higher-energy doublon states. Additionally, we provide the full quasi-independent pair spectrum in the Supplementary. 

To explain why the doublon band is actually formed while there are no matrix elements in $\hat H$ connecting the doublon states, one has to assume that at strong repulsion/attraction $U$, a pair tunnels through virtual splits. To second order, its hopping is $2t_j^2/U$ and its on-site energy is $U+2(t_{j-1}^2+t_j^2)/U$, with $t_0=t_N=0$ at the open ends~\cite{gorlach2017}. This means that the Fibonacci bonds determine both pair hopping and local energy shifts, including a reduced shift at each end; the latter are absent from the one-boson Hamiltonian, so its restriction on right-edge states does not carry over to doublons. An example worth mentioning regarding the mechanism is an exact angle-independent doublon state at $E=U$ with uniform same-site probability, $P_d=1$ and $I_2=1/N$.  It appears since the tunneling term on bond $j$ connects both $\ket{j,j}$ and $\ket{j+1, j+1}$ to $\ket{j,j+1}$ with the same matrix element $\sqrt2\,t_j$. Their contributions therefore cancel out for an eigenstate with alternating amplitudes:
\begin{equation}
		\ket{\eta}=\frac{1}{\sqrt N}\sum_{j=1}^{N}(-1)^{j-1}\ket{j,j},\quad \hat H\ket{\eta}=U\ket{\eta}.
	\label{eq:eta_pair}
\end{equation}
The cancellation occurs for any Fibonacci word at every $\theta$, and provides a nice analytical benchmark. This state has the lowest energy in the doublon band, and produces a flat line in Fig. \ref{fig:doublon_spectra}. Interestingly, the adjacent higher-energy states have profiles $\left<j,j|\psi_k\right>$ with the same high-frequency alternating structure, but additionally with an approximately harmonic modulation despite the quasicrystal disorder. We provide a figure illustrating the limits of this second-order model with respect to the parameter $U$, and the plots of the lowest doublon profiles in the Supplementary.

Fig.~\ref{fig:doublon_states} compares several two-boson profiles at $N=56$ including both doublons and separated pairs.  For smoother visualizations, we use here a negative hopping Hamiltonian compared to Eq.~\eqref{eq:bh}; the separated-pair solutions are related by the gauge change $\psi_{ij}\mapsto(-1)^{i+j}\psi_{ij}$, which preserves energies and probabilities (the same-site coefficients $\psi_{jj}$ are unchanged).  The bulk independent-pair example is shown in Fig.~\ref{fig:doublon_states}(a) and has $I_2\simeq2.1\cdot10^{-3}$. The seven independent-pair profiles (b--h)  have much larger $I_2\approx 0.06$--$0.24$, with minimal double occupancy $P_d<3\%$. As on can see, their peaks occur off the diagonal but still concentrate both particles near the same end. In contrast, the doublon profiles (i--l) have $P_d>90\%$ and are dominated by the diagonal. We observe that both types of two-boson states occur both at the left and right boundaries, in contrast to the single-boson subspace.

By matching the edge-aligned same-site amplitudes, one can identify the four confined doublons $\alpha$--$\delta$ in Fig.~\ref{fig:doublon_states}(i--l) with their twin states at $N=145$ ($k_{\rm d}=111,56,90,89$), respectively. Their fitted amplitude decay lengths are annotated in Fig.~\ref{fig:doublon_spectra}. For the state $\alpha$, the localization length fits give slightly different values for $N=90$ and $145$; the latter case is closer to convergence. In contrast, the highly localized right-edge doublon labeled $\gamma$ is much more stable; in the range of $S_{8-11}$, its energy agrees to six decimal places. Generally, the right-edge doublons recur in adjacent generations, unlike the left-edge one-boson states whose persistence follows the left termination. All found two-boson edge states, nevertheless, are expected to occur indefinitely at arbitrarily large $N$ being an inherent attribute of the FBH model.

In conclusion, we have studied the Fibonacci-Bose-Hubbard clusters in the one- and two-particle sectors, and suggest that they could be promising for experimental quantum simulations beyond classical tractability. The combination of simply defined physical structure, reproducible boundary profiles, hierarchical spectra, and an exact zero-energy state makes this type of Hamiltonian a rich and interesting research object beyond uniform and alternating periodic chains. The band structure can be directly measured using transmission spectroscopy, as in recent experiments \cite{fedorov2021, jouanny2025high}.

Present eigenstate analysis can be connected to transport properties by studying the response of the clusters to a boundary drive. In these experiments, many-body dynamics will be revealed by including additional multi-boson sectors in the truncated Hamiltonian, which we actually expect to be quite demanding in terms of computation. For the complete words considered here, directly observable new phenomena that one can expect in these experiments include the in-gap states in both the single and several-photon bands, intra- and interband multi-photon transitions, cross-Kerr interactions between the different subspaces when a bichromatic excitation is applied, etc. Additionally, we note that using tunable couplers in an experimental setting will allow one to reproduce the full spectral plots presented here, and monitor how the eigenstates vary along the way, including measuring the localization length, which is a great quality benchmark for any experimental platform. Another interesting direction of experiments is to vary the on-site interaction potential from chain to chain, observing the breakdown of the second-order models. Finally, the most interesting and involved experiments in this context will likely concern quasi-periodic arrays beyond one dimension.

\begin{acknowledgments}

The Python code for the numerical simulations and the \textit{matplotlib} figures was drafted via OpenAI Astra and finalized by hand. The authors thank Ivan Berkut, Vladimir Zhaketov, Maxim Gorlach, each other, Editors, anonymous Referees.

\end{acknowledgments}

\bibliographystyle{apsrev4-2-threeauthors}
\bibliography{references}

\begin{thebibliography}{23}%
\makeatletter
\providecommand \@ifxundefined [1]{%
 \@ifx{#1\undefined}
}%
\providecommand \@ifnum [1]{%
 \ifnum #1\expandafter \@firstoftwo
 \else \expandafter \@secondoftwo
 \fi
}%
\providecommand \@ifx [1]{%
 \ifx #1\expandafter \@firstoftwo
 \else \expandafter \@secondoftwo
 \fi
}%
\providecommand \natexlab [1]{#1}%
\providecommand \enquote  [1]{``#1''}%
\providecommand \bibnamefont  [1]{#1}%
\providecommand \bibfnamefont [1]{#1}%
\providecommand \citenamefont [1]{#1}%
\providecommand \href@noop [0]{\@secondoftwo}%
\providecommand \href [0]{\begingroup \@sanitize@url \@href}%
\providecommand \@href[1]{\@@startlink{#1}\@@href}%
\providecommand \@@href[1]{\endgroup#1\@@endlink}%
\providecommand \@sanitize@url [0]{\catcode `\\12\catcode `\$12\catcode
  `\&12\catcode `\#12\catcode `\^12\catcode `\_12\catcode `\%12\relax}%
\providecommand \@@startlink[1]{}%
\providecommand \@@endlink[0]{}%
\providecommand \url  [0]{\begingroup\@sanitize@url \@url }%
\providecommand \@url [1]{\endgroup\@href {#1}{\urlprefix }}%
\providecommand \urlprefix  [0]{URL }%
\providecommand \Eprint [0]{\href }%
\providecommand \doibase [0]{https://doi.org/}%
\providecommand \selectlanguage [0]{\@gobble}%
\providecommand \bibinfo  [0]{\@secondoftwo}%
\providecommand \bibfield  [0]{\@secondoftwo}%
\providecommand \translation [1]{[#1]}%
\providecommand \BibitemOpen [0]{}%
\providecommand \bibitemStop [0]{}%
\providecommand \bibitemNoStop [0]{.\EOS\space}%
\providecommand \EOS [0]{\spacefactor3000\relax}%
\providecommand \BibitemShut  [1]{\csname bibitem#1\endcsname}%
\let\auto@bib@innerbib\@empty
\bibitem [{\citenamefont {Kolovsky}(2016)}]{kolovsky2016}%
  \BibitemOpen
  \bibfield  {author} {\bibinfo {author} {\bibfnamefont {A.~R.}\ \bibnamefont
  {Kolovsky}},\ }\href {https://doi.org/10.1142/S0217979216300097} {\bibfield
  {journal} {\bibinfo  {journal} {International Journal of Modern Physics B}\
  }\textbf {\bibinfo {volume} {30}},\ \bibinfo {pages} {1630009} (\bibinfo
  {year} {2016})}\BibitemShut {NoStop}%
\bibitem [{\citenamefont {Jaksch}\ \emph {et~al.}(1998)\citenamefont {Jaksch},
  \citenamefont {Bruder}, \citenamefont {Cirac}, \citenamefont {Gardiner},\
  and\ \citenamefont {Zoller}}]{jaksch1998}%
  \BibitemOpen
  \bibfield  {author} {\bibinfo {author} {\bibfnamefont {D.}~\bibnamefont
  {Jaksch}}, \bibinfo {author} {\bibfnamefont {C.}~\bibnamefont {Bruder}},
  \bibinfo {author} {\bibfnamefont {J.~I.}\ \bibnamefont {Cirac}}, \emph
  {et~al.},\ }\href {https://doi.org/10.1103/PhysRevLett.81.3108} {\bibfield
  {journal} {\bibinfo  {journal} {Physical Review Letters}\ }\textbf {\bibinfo
  {volume} {81}},\ \bibinfo {pages} {3108} (\bibinfo {year}
  {1998})}\BibitemShut {NoStop}%
\bibitem [{\citenamefont {Kaufman}\ \emph {et~al.}(2016)\citenamefont
  {Kaufman}, \citenamefont {Tai}, \citenamefont {Lukin}, \citenamefont
  {Rispoli}, \citenamefont {Schittko}, \citenamefont {Preiss},\ and\
  \citenamefont {Greiner}}]{kaufman2016}%
  \BibitemOpen
  \bibfield  {author} {\bibinfo {author} {\bibfnamefont {A.~M.}\ \bibnamefont
  {Kaufman}}, \bibinfo {author} {\bibfnamefont {M.~E.}\ \bibnamefont {Tai}},
  \bibinfo {author} {\bibfnamefont {A.}~\bibnamefont {Lukin}}, \emph {et~al.},\
  }\href {https://doi.org/10.1126/science.aaf6725} {\bibfield  {journal}
  {\bibinfo  {journal} {Science}\ }\textbf {\bibinfo {volume} {353}},\ \bibinfo
  {pages} {794} (\bibinfo {year} {2016})}\BibitemShut {NoStop}%
\bibitem [{\citenamefont {Roushan}\ \emph {et~al.}(2017)\citenamefont
  {Roushan}, \citenamefont {Neill}, \citenamefont {Tangpanitanon},
  \citenamefont {Bastidas}, \citenamefont {Megrant}, \citenamefont {Barends},
  \citenamefont {Chen}, \citenamefont {Chen}, \citenamefont {Chiaro},
  \citenamefont {Dunsworth}, \citenamefont {Fowler}, \citenamefont {Foxen},
  \citenamefont {Giustina}, \citenamefont {Jeffrey}, \citenamefont {Kelly},
  \citenamefont {Lucero}, \citenamefont {Mutus}, \citenamefont {Neeley},
  \citenamefont {Quintana}, \citenamefont {Sank}, \citenamefont {Vainsencher},
  \citenamefont {Wenner}, \citenamefont {White}, \citenamefont {Neven},
  \citenamefont {Angelakis},\ and\ \citenamefont {Martinis}}]{roushan2017}%
  \BibitemOpen
  \bibfield  {author} {\bibinfo {author} {\bibfnamefont {P.}~\bibnamefont
  {Roushan}}, \bibinfo {author} {\bibfnamefont {C.}~\bibnamefont {Neill}},
  \bibinfo {author} {\bibfnamefont {J.}~\bibnamefont {Tangpanitanon}}, \emph
  {et~al.},\ }\href {https://doi.org/10.1126/science.aao1401} {\bibfield
  {journal} {\bibinfo  {journal} {Science}\ }\textbf {\bibinfo {volume}
  {358}},\ \bibinfo {pages} {1175} (\bibinfo {year} {2017})}\BibitemShut
  {NoStop}%
\bibitem [{\citenamefont {Fedorov}\ \emph {et~al.}(2021)\citenamefont
  {Fedorov}, \citenamefont {Remizov}, \citenamefont {Shapiro}, \citenamefont
  {Pogosov}, \citenamefont {Egorova}, \citenamefont {Tsitsilin}, \citenamefont
  {Andronik}, \citenamefont {Dobronosova}, \citenamefont {Rodionov},
  \citenamefont {Astafiev},\ and\ \citenamefont {Ustinov}}]{fedorov2021}%
  \BibitemOpen
  \bibfield  {author} {\bibinfo {author} {\bibfnamefont {G.~P.}\ \bibnamefont
  {Fedorov}}, \bibinfo {author} {\bibfnamefont {S.~V.}\ \bibnamefont
  {Remizov}}, \bibinfo {author} {\bibfnamefont {D.~S.}\ \bibnamefont
  {Shapiro}}, \emph {et~al.},\ }\href
  {https://doi.org/10.1103/PhysRevLett.126.180503} {\bibfield  {journal}
  {\bibinfo  {journal} {Physical Review Letters}\ }\textbf {\bibinfo {volume}
  {126}},\ \bibinfo {pages} {180503} (\bibinfo {year} {2021})}\BibitemShut
  {NoStop}%
\bibitem [{\citenamefont {Wang}\ \emph {et~al.}(2025)\citenamefont {Wang},
  \citenamefont {Parker}, \citenamefont {Champion},\ and\ \citenamefont
  {Blok}}]{wang2025}%
  \BibitemOpen
  \bibfield  {author} {\bibinfo {author} {\bibfnamefont {Z.}~\bibnamefont
  {Wang}}, \bibinfo {author} {\bibfnamefont {R.~W.}\ \bibnamefont {Parker}},
  \bibinfo {author} {\bibfnamefont {E.}~\bibnamefont {Champion}}, \emph
  {et~al.},\ }\href {https://doi.org/10.1103/PhysRevApplied.23.034046}
  {\bibfield  {journal} {\bibinfo  {journal} {Physical Review Applied}\
  }\textbf {\bibinfo {volume} {23}},\ \bibinfo {pages} {034046} (\bibinfo
  {year} {2025})}\BibitemShut {NoStop}%
\bibitem [{\citenamefont {Besedin}\ \emph {et~al.}(2021)\citenamefont
  {Besedin}, \citenamefont {Gorlach}, \citenamefont {Abramov}, \citenamefont
  {Tsitsilin}, \citenamefont {Moskalenko}, \citenamefont {Dobronosova},
  \citenamefont {Moskalev}, \citenamefont {Matanin}, \citenamefont {Smirnov},
  \citenamefont {Rodionov}, \citenamefont {Poddubny},\ and\ \citenamefont
  {Ustinov}}]{besedin2021}%
  \BibitemOpen
  \bibfield  {author} {\bibinfo {author} {\bibfnamefont {I.~S.}\ \bibnamefont
  {Besedin}}, \bibinfo {author} {\bibfnamefont {M.~A.}\ \bibnamefont
  {Gorlach}}, \bibinfo {author} {\bibfnamefont {N.~N.}\ \bibnamefont
  {Abramov}}, \emph {et~al.},\ }\href
  {https://doi.org/10.1103/PhysRevB.103.224520} {\bibfield  {journal} {\bibinfo
   {journal} {Physical Review B}\ }\textbf {\bibinfo {volume} {103}},\ \bibinfo
  {pages} {224520} (\bibinfo {year} {2021})}\BibitemShut {NoStop}%
\bibitem [{\citenamefont {Konopleva}\ \emph
  {et~al.}(2026{\natexlab{a}})\citenamefont {Konopleva}, \citenamefont
  {Fedorov},\ and\ \citenamefont {Astafiev}}]{konopleva2026new}%
  \BibitemOpen
  \bibfield  {author} {\bibinfo {author} {\bibfnamefont {E.~E.}\ \bibnamefont
  {Konopleva}}, \bibinfo {author} {\bibfnamefont {G.~P.}\ \bibnamefont
  {Fedorov}},\ and\ \bibinfo {author} {\bibfnamefont {O.~V.}\ \bibnamefont
  {Astafiev}},\ }\href@noop {} {\bibfield  {journal} {\bibinfo  {journal}
  {Radiophysics and Quantum Electronics}\ ,\ \bibinfo {pages} {1}} (\bibinfo
  {year} {2026}{\natexlab{a}})}\BibitemShut {NoStop}%
\bibitem [{\citenamefont {Konopleva}\ \emph
  {et~al.}(2026{\natexlab{b}})\citenamefont {Konopleva}, \citenamefont
  {Fedorov},\ and\ \citenamefont {Astafiev}}]{konopleva2026nontrivial}%
  \BibitemOpen
  \bibfield  {author} {\bibinfo {author} {\bibfnamefont {E.~E.}\ \bibnamefont
  {Konopleva}}, \bibinfo {author} {\bibfnamefont {G.~P.}\ \bibnamefont
  {Fedorov}},\ and\ \bibinfo {author} {\bibfnamefont {O.~V.}\ \bibnamefont
  {Astafiev}},\ }\href@noop {} {\bibfield  {journal} {\bibinfo  {journal}
  {Advanced Quantum Technologies}\ }\textbf {\bibinfo {volume} {9}},\ \bibinfo
  {pages} {e00912} (\bibinfo {year} {2026}{\natexlab{b}})}\BibitemShut
  {NoStop}%
\bibitem [{\citenamefont {Sandberg}\ \emph {et~al.}(2024)\citenamefont
  {Sandberg}, \citenamefont {Awoga}, \citenamefont {Black-Schaffer},\ and\
  \citenamefont {Holmvall}}]{sandberg2024josephson}%
  \BibitemOpen
  \bibfield  {author} {\bibinfo {author} {\bibfnamefont {A.}~\bibnamefont
  {Sandberg}}, \bibinfo {author} {\bibfnamefont {O.~A.}\ \bibnamefont {Awoga}},
  \bibinfo {author} {\bibfnamefont {A.~M.}\ \bibnamefont {Black-Schaffer}},
  \emph {et~al.},\ }\href@noop {} {\bibfield  {journal} {\bibinfo  {journal}
  {Physical Review B}\ }\textbf {\bibinfo {volume} {110}},\ \bibinfo {pages}
  {104513} (\bibinfo {year} {2024})}\BibitemShut {NoStop}%
\bibitem [{\citenamefont {Su}\ \emph {et~al.}(1979)\citenamefont {Su},
  \citenamefont {Schrieffer},\ and\ \citenamefont {Heeger}}]{ssh1979}%
  \BibitemOpen
  \bibfield  {author} {\bibinfo {author} {\bibfnamefont {W.~P.}\ \bibnamefont
  {Su}}, \bibinfo {author} {\bibfnamefont {J.~R.}\ \bibnamefont {Schrieffer}},\
  and\ \bibinfo {author} {\bibfnamefont {A.~J.}\ \bibnamefont {Heeger}},\
  }\href {https://doi.org/10.1103/PhysRevLett.42.1698} {\bibfield  {journal}
  {\bibinfo  {journal} {Physical Review Letters}\ }\textbf {\bibinfo {volume}
  {42}},\ \bibinfo {pages} {1698} (\bibinfo {year} {1979})}\BibitemShut
  {NoStop}%
\bibitem [{\citenamefont {Kohmoto}\ \emph {et~al.}(1983)\citenamefont
  {Kohmoto}, \citenamefont {Kadanoff},\ and\ \citenamefont
  {Tang}}]{kohmoto1983}%
  \BibitemOpen
  \bibfield  {author} {\bibinfo {author} {\bibfnamefont {M.}~\bibnamefont
  {Kohmoto}}, \bibinfo {author} {\bibfnamefont {L.~P.}\ \bibnamefont
  {Kadanoff}},\ and\ \bibinfo {author} {\bibfnamefont {C.}~\bibnamefont
  {Tang}},\ }\href {https://doi.org/10.1103/PhysRevLett.50.1870} {\bibfield
  {journal} {\bibinfo  {journal} {Physical Review Letters}\ }\textbf {\bibinfo
  {volume} {50}},\ \bibinfo {pages} {1870} (\bibinfo {year}
  {1983})}\BibitemShut {NoStop}%
\bibitem [{\citenamefont {Kohmoto}\ \emph {et~al.}(1987)\citenamefont
  {Kohmoto}, \citenamefont {Sutherland},\ and\ \citenamefont
  {Tang}}]{kohmoto1987}%
  \BibitemOpen
  \bibfield  {author} {\bibinfo {author} {\bibfnamefont {M.}~\bibnamefont
  {Kohmoto}}, \bibinfo {author} {\bibfnamefont {B.}~\bibnamefont
  {Sutherland}},\ and\ \bibinfo {author} {\bibfnamefont {C.}~\bibnamefont
  {Tang}},\ }\href {https://doi.org/10.1103/PhysRevB.35.1020} {\bibfield
  {journal} {\bibinfo  {journal} {Physical Review B}\ }\textbf {\bibinfo
  {volume} {35}},\ \bibinfo {pages} {1020} (\bibinfo {year}
  {1987})}\BibitemShut {NoStop}%
\bibitem [{\citenamefont {Jagannathan}(2021)}]{jagannathan2021}%
  \BibitemOpen
  \bibfield  {author} {\bibinfo {author} {\bibfnamefont {A.}~\bibnamefont
  {Jagannathan}},\ }\href {https://doi.org/10.1103/RevModPhys.93.045001}
  {\bibfield  {journal} {\bibinfo  {journal} {Reviews of Modern Physics}\
  }\textbf {\bibinfo {volume} {93}},\ \bibinfo {pages} {045001} (\bibinfo
  {year} {2021})}\BibitemShut {NoStop}%
\bibitem [{\citenamefont {Mac{\'e}}\ \emph {et~al.}(2017)\citenamefont
  {Mac{\'e}}, \citenamefont {Jagannathan}, \citenamefont {Kalugin},
  \citenamefont {Mosseri},\ and\ \citenamefont {Pi{\'e}chon}}]{mace2017}%
  \BibitemOpen
  \bibfield  {author} {\bibinfo {author} {\bibfnamefont {N.}~\bibnamefont
  {Mac{\'e}}}, \bibinfo {author} {\bibfnamefont {A.}~\bibnamefont
  {Jagannathan}}, \bibinfo {author} {\bibfnamefont {P.}~\bibnamefont
  {Kalugin}}, \emph {et~al.},\ }\href
  {https://doi.org/10.1103/PhysRevB.96.045138} {\bibfield  {journal} {\bibinfo
  {journal} {Physical Review B}\ }\textbf {\bibinfo {volume} {96}},\ \bibinfo
  {pages} {045138} (\bibinfo {year} {2017})}\BibitemShut {NoStop}%
\bibitem [{\citenamefont {Zijlstra}\ \emph {et~al.}(1999)\citenamefont
  {Zijlstra}, \citenamefont {Fasolino},\ and\ \citenamefont
  {Janssen}}]{zijlstra1999}%
  \BibitemOpen
  \bibfield  {author} {\bibinfo {author} {\bibfnamefont {E.~S.}\ \bibnamefont
  {Zijlstra}}, \bibinfo {author} {\bibfnamefont {A.}~\bibnamefont {Fasolino}},\
  and\ \bibinfo {author} {\bibfnamefont {T.}~\bibnamefont {Janssen}},\ }\href
  {https://doi.org/10.1103/PhysRevB.59.302} {\bibfield  {journal} {\bibinfo
  {journal} {Physical Review B}\ }\textbf {\bibinfo {volume} {59}},\ \bibinfo
  {pages} {302} (\bibinfo {year} {1999})}\BibitemShut {NoStop}%
\bibitem [{\citenamefont {Gorlach}\ and\ \citenamefont
  {Poddubny}(2017)}]{gorlach2017}%
  \BibitemOpen
  \bibfield  {author} {\bibinfo {author} {\bibfnamefont {M.~A.}\ \bibnamefont
  {Gorlach}}\ and\ \bibinfo {author} {\bibfnamefont {A.~N.}\ \bibnamefont
  {Poddubny}},\ }\href {https://doi.org/10.1103/PhysRevA.95.053866} {\bibfield
  {journal} {\bibinfo  {journal} {Physical Review A}\ }\textbf {\bibinfo
  {volume} {95}},\ \bibinfo {pages} {053866} (\bibinfo {year}
  {2017})}\BibitemShut {NoStop}%
\bibitem [{\citenamefont {Di~Liberto}\ \emph {et~al.}(2016)\citenamefont
  {Di~Liberto}, \citenamefont {Recati}, \citenamefont {Carusotto},\ and\
  \citenamefont {Menotti}}]{diliberto2016}%
  \BibitemOpen
  \bibfield  {author} {\bibinfo {author} {\bibfnamefont {M.}~\bibnamefont
  {Di~Liberto}}, \bibinfo {author} {\bibfnamefont {A.}~\bibnamefont {Recati}},
  \bibinfo {author} {\bibfnamefont {I.}~\bibnamefont {Carusotto}}, \emph
  {et~al.},\ }\href {https://doi.org/10.1103/PhysRevA.94.062704} {\bibfield
  {journal} {\bibinfo  {journal} {Physical Review A}\ }\textbf {\bibinfo
  {volume} {94}},\ \bibinfo {pages} {062704} (\bibinfo {year}
  {2016})}\BibitemShut {NoStop}%
\bibitem [{\citenamefont {Damanik}\ and\ \citenamefont
  {Gorodetski}(2008)}]{damanik2008spectrum}%
  \BibitemOpen
  \bibfield  {author} {\bibinfo {author} {\bibfnamefont {D.}~\bibnamefont
  {Damanik}}\ and\ \bibinfo {author} {\bibfnamefont {A.}~\bibnamefont
  {Gorodetski}},\ }\href@noop {} {\bibfield  {journal} {\bibinfo  {journal}
  {arXiv:0807.3024}\ } (\bibinfo {year} {2008})}\BibitemShut {NoStop}%
\bibitem [{\citenamefont {Valiente}\ and\ \citenamefont
  {Petrosyan}(2008)}]{valiente2008}%
  \BibitemOpen
  \bibfield  {author} {\bibinfo {author} {\bibfnamefont {M.}~\bibnamefont
  {Valiente}}\ and\ \bibinfo {author} {\bibfnamefont {D.}~\bibnamefont
  {Petrosyan}},\ }\href {https://doi.org/10.1088/0953-4075/41/16/161002}
  {\bibfield  {journal} {\bibinfo  {journal} {Journal of Physics B: Atomic,
  Molecular and Optical Physics}\ }\textbf {\bibinfo {volume} {41}},\ \bibinfo
  {pages} {161002} (\bibinfo {year} {2008})}\BibitemShut {NoStop}%
\bibitem [{\citenamefont {Pinto}\ \emph {et~al.}(2009)\citenamefont {Pinto},
  \citenamefont {Haque},\ and\ \citenamefont {Flach}}]{pinto2009}%
  \BibitemOpen
  \bibfield  {author} {\bibinfo {author} {\bibfnamefont {R.~A.}\ \bibnamefont
  {Pinto}}, \bibinfo {author} {\bibfnamefont {M.}~\bibnamefont {Haque}},\ and\
  \bibinfo {author} {\bibfnamefont {S.}~\bibnamefont {Flach}},\ }\href
  {https://doi.org/10.1103/PhysRevA.79.052118} {\bibfield  {journal} {\bibinfo
  {journal} {Physical Review A}\ }\textbf {\bibinfo {volume} {79}},\ \bibinfo
  {pages} {052118} (\bibinfo {year} {2009})}\BibitemShut {NoStop}%
\bibitem [{\citenamefont {Di~Liberto}\ \emph {et~al.}(2017)\citenamefont
  {Di~Liberto}, \citenamefont {Recati}, \citenamefont {Carusotto},\ and\
  \citenamefont {Menotti}}]{diliberto2017}%
  \BibitemOpen
  \bibfield  {author} {\bibinfo {author} {\bibfnamefont {M.}~\bibnamefont
  {Di~Liberto}}, \bibinfo {author} {\bibfnamefont {A.}~\bibnamefont {Recati}},
  \bibinfo {author} {\bibfnamefont {I.}~\bibnamefont {Carusotto}}, \emph
  {et~al.},\ }\href {https://doi.org/10.1140/epjst/e2016-60388-y} {\bibfield
  {journal} {\bibinfo  {journal} {European Physical Journal Special Topics}\
  }\textbf {\bibinfo {volume} {226}},\ \bibinfo {pages} {2751} (\bibinfo {year}
  {2017})}\BibitemShut {NoStop}%
\bibitem [{\citenamefont {Jouanny}\ \emph {et~al.}(2025)\citenamefont
  {Jouanny}, \citenamefont {Frasca}, \citenamefont {Weibel}, \citenamefont
  {Peyruchat}, \citenamefont {Scigliuzzo}, \citenamefont {Oppliger},
  \citenamefont {De~Palma}, \citenamefont {Sbroggi{\`o}}, \citenamefont
  {Beaulieu}, \citenamefont {Zilberberg} \emph {et~al.}}]{jouanny2025high}%
  \BibitemOpen
  \bibfield  {author} {\bibinfo {author} {\bibfnamefont {V.}~\bibnamefont
  {Jouanny}}, \bibinfo {author} {\bibfnamefont {S.}~\bibnamefont {Frasca}},
  \bibinfo {author} {\bibfnamefont {V.~J.}\ \bibnamefont {Weibel}}, \emph
  {et~al.},\ }\href@noop {} {\bibfield  {journal} {\bibinfo  {journal} {Nature
  Communications}\ }\textbf {\bibinfo {volume} {16}},\ \bibinfo {pages} {3396}
  (\bibinfo {year} {2025})}\BibitemShut {NoStop}%
\end{thebibliography}%


\begin{thebibliography}{1}

\bibitem{kraus2012}
Yaacov~E. Kraus and Oded Zilberberg.
\newblock Topological equivalence between the {Fibonacci} quasicrystal and the
  {Harper} model.
\newblock {\em Physical Review Letters}, 109:116404, 2012.

\bibitem{verbin2015}
Mor Verbin, Oded Zilberberg, Yoav Lahini, Yaacov~E. Kraus, and Yaron
  Silberberg.
\newblock Topological pumping over a photonic {Fibonacci} quasicrystal.
\newblock {\em Physical Review B}, 91:064201, 2015.

\bibitem{kohmoto1983}
Mahito Kohmoto, Leo~P. Kadanoff, and Chao Tang.
\newblock Localization problem in one dimension: Mapping and escape.
\newblock {\em Physical Review Letters}, 50:1870, 1983.

\bibitem{jagannathan2021}
Anuradha Jagannathan.
\newblock The {Fibonacci} quasicrystal: Case study of hidden dimensions and
  multifractality.
\newblock {\em Reviews of Modern Physics}, 93:045001, 2021.

\end{thebibliography}
\end{document}


\maketitle
\vspace{-2em}

\begin{figure}[htbp]
\centering
\includegraphics[width=\textwidth,height=.62\textheight,keepaspectratio]{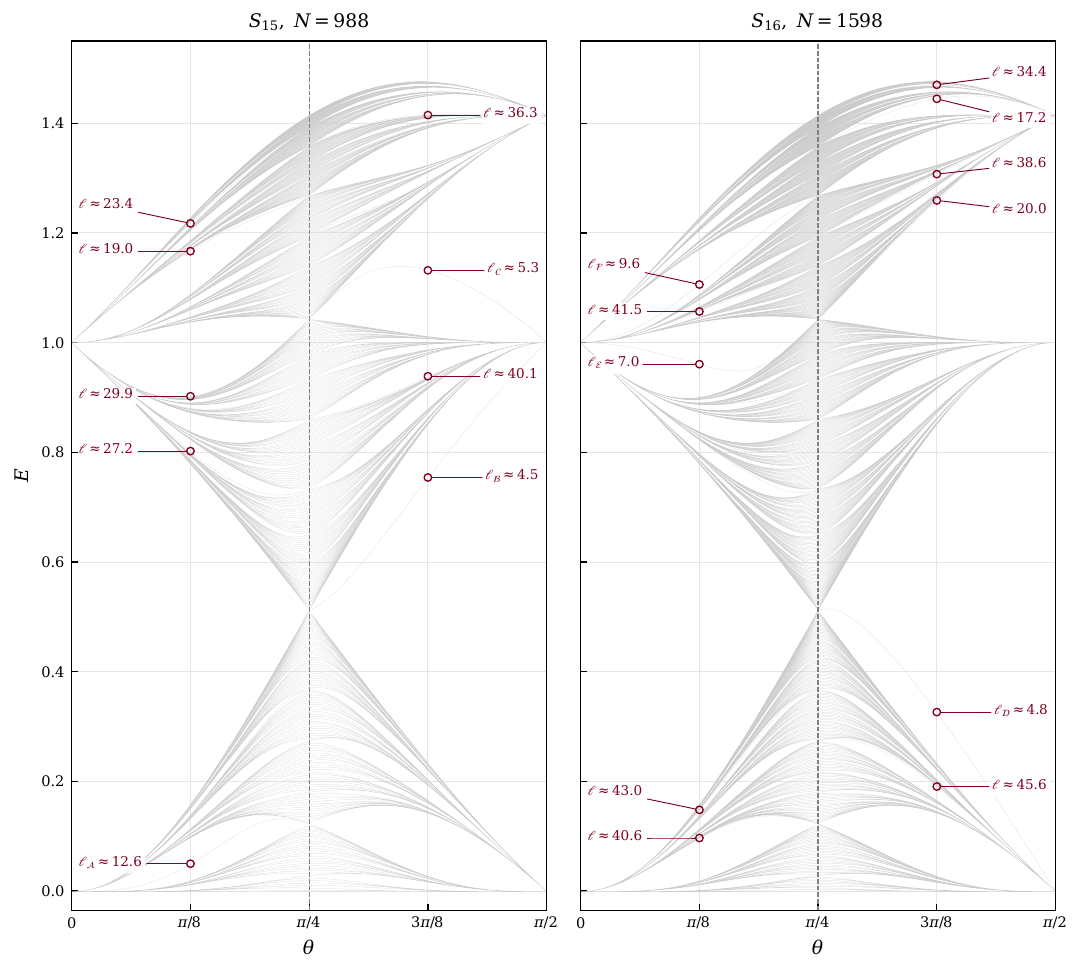}
\caption{Nonnegative one-boson spectra for $S_{15}$ ($N=988$) and $S_{16}$ ($N=1598$), with $t_A=\cos\theta$ and $t_B=\sin\theta$. Circles mark left-edge states at $\theta=\pi/8$ and $3\pi/8$ with fitted amplitude decay length $\ell<50$ sites. The lengths are obtained by linear fits to the logarithms of local maxima of $|\psi_j|$, using the same fits as in main-text Fig.~2. Subscripts $\mathcal A$--$\mathcal F$ identify the corresponding branches. The dashed line marks the uniform chain at $\theta=\pi/4$.}
\label{fig:whisker_overview}
\end{figure}
\clearpage

\begin{figure}[htbp]
\centering
\includegraphics[page=1,width=\textwidth]{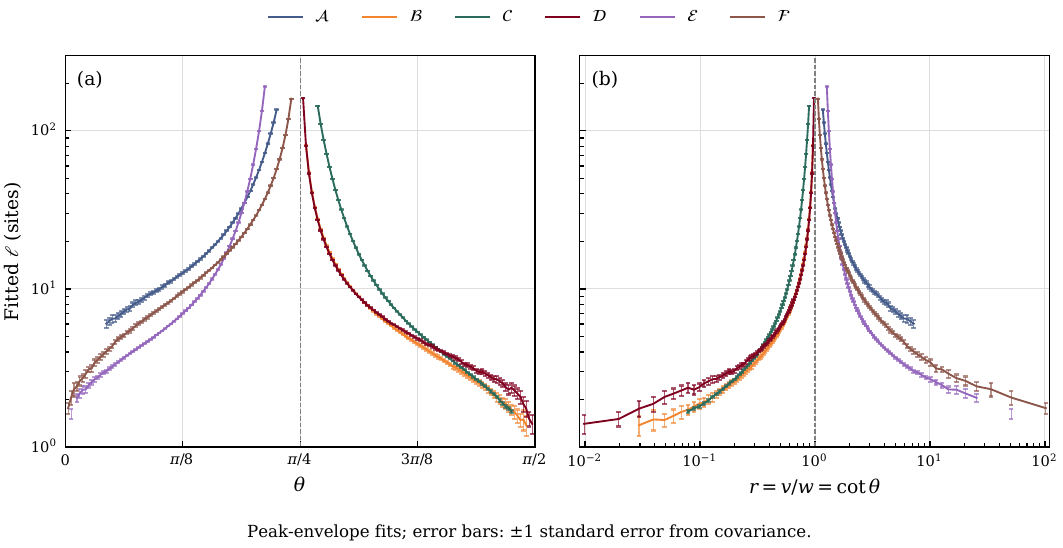}

\includegraphics[page=2,width=\textwidth]{figures/fig_localization_envelope_fits.pdf}
\caption{Amplitude decay lengths of the six left-edge branches $\mathcal A$--$\mathcal F$ versus (a) $\theta$ and (b) $v/w=\cot\theta$. The branches are continued from the states marked in Fig.~\ref{fig:whisker_overview}, using larger chains with $N=2585$ for $\mathcal A$--$\mathcal C$ and $N=4182$ for $\mathcal D$--$\mathcal F$. Fits use $\log|\psi_{j_p}|=\log C-(j_p-j_0)/\ell$ at local amplitude maxima $j_p$, where $j_0$ is the largest peak's position}
\label{fig:envelope_scan}
\end{figure}

\clearpage

\begin{figure}[htbp]
\centering
\includegraphics[width=\textwidth]{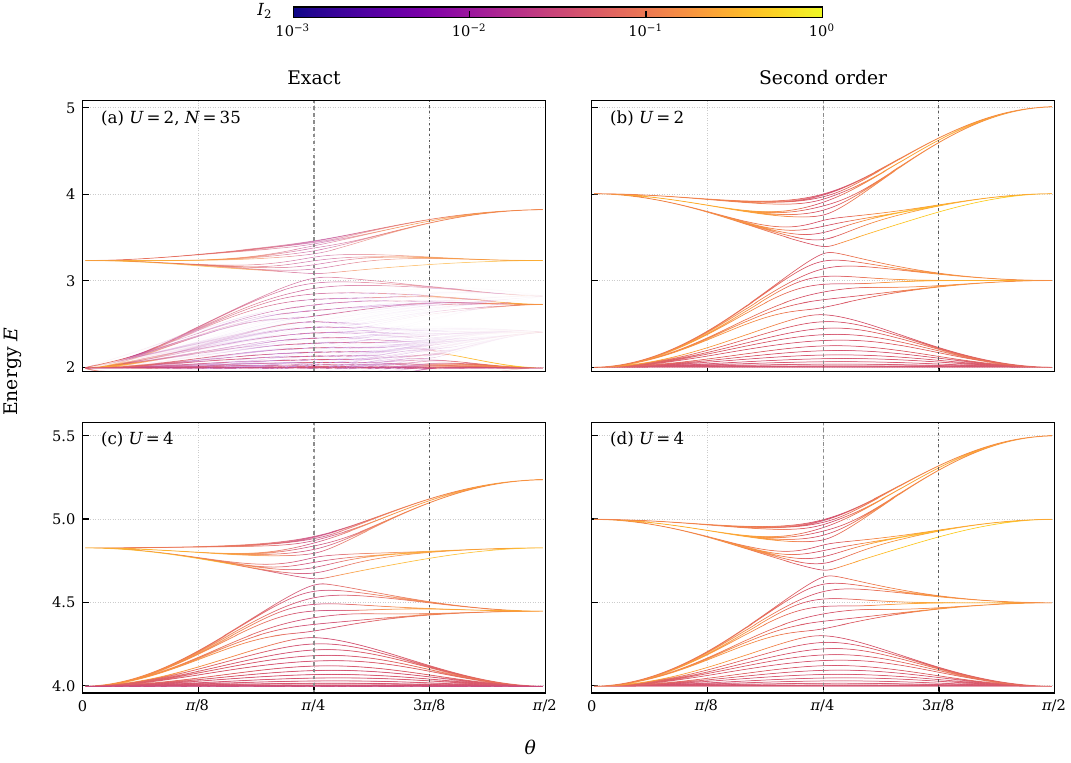}
\caption{Exact (left) and second-order (right) two-boson spectra for $S_8$, $N=35$, at $U=2$ (top) and $U=4$ (bottom). The effective pair Hamiltonian has hopping $2t_j^2/U$ and diagonal energy $U+2(t_{j-1}^2+t_j^2)/U$, with $t_0=t_N=0$. All 630 exact levels are retained in the calculation; the displayed energy window focuses on the pair band. Exact lines become fainter and thinner as their same-site probability $P_d$ decreases. Colors give the full-state $I_2=\sum_{i\leq j}|\langle i,j|\psi\rangle|^4$ for the exact states. For the effective model, the normalized pair amplitudes are embedded in the same-site configurations, with all split-pair amplitudes zero. Both columns share the physical logarithmic color scale and the energy limits within each row. The sweep uses 159 interior angles. Values below $10^{-3}$ are clipped on the color scale. Near unresolved degeneracies, individual-state colors can depend on the numerical eigenbasis.}
\label{fig:doublon_comparison}
\end{figure}
\clearpage

\begin{figure}[htbp]
\centering
\includegraphics[width=.9\textwidth,height=.76\textheight,keepaspectratio]{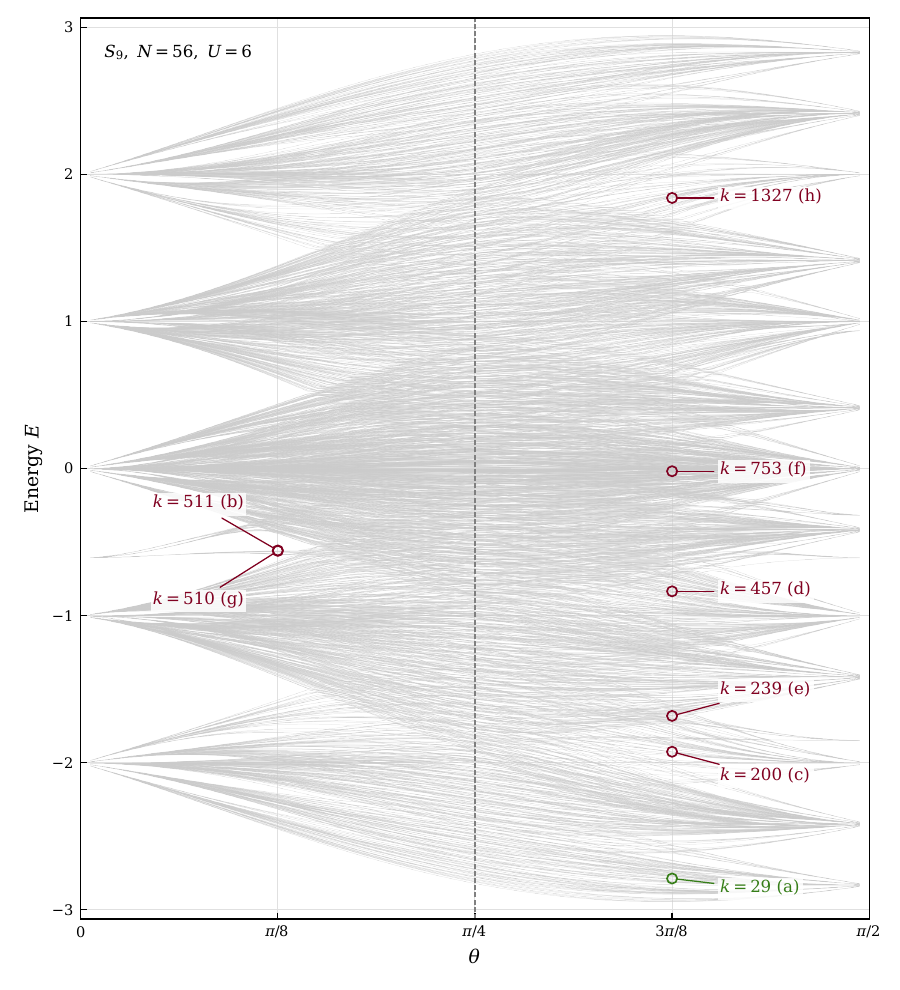}
\caption{Split-pair sector of the exact $S_9$ ($N=56$) two-boson spectrum at $U=6$. Gray lines show the lower 1540 eigenvalues; the upper 56 doublon states are omitted. Circles at $\theta=\pi/8$ and $3\pi/8$ mark the seven edge-concentrated profiles in main-text Fig.~5(b--h), together with its bulk example (a), marked in green. Labels give the zero-based full-spectrum index $k$ and the corresponding main-text panel. The nearby levels $k=510,511$ have opposite-edge profiles. The sweep contains 79 interior angles and uses negative hopping amplitudes, related to the main-text convention by a gauge transformation that preserves the energies.}
\label{fig:split_pair_spectrum}
\end{figure}
\clearpage

\begin{figure}[htbp]
\centering
\includegraphics[width=\textwidth]{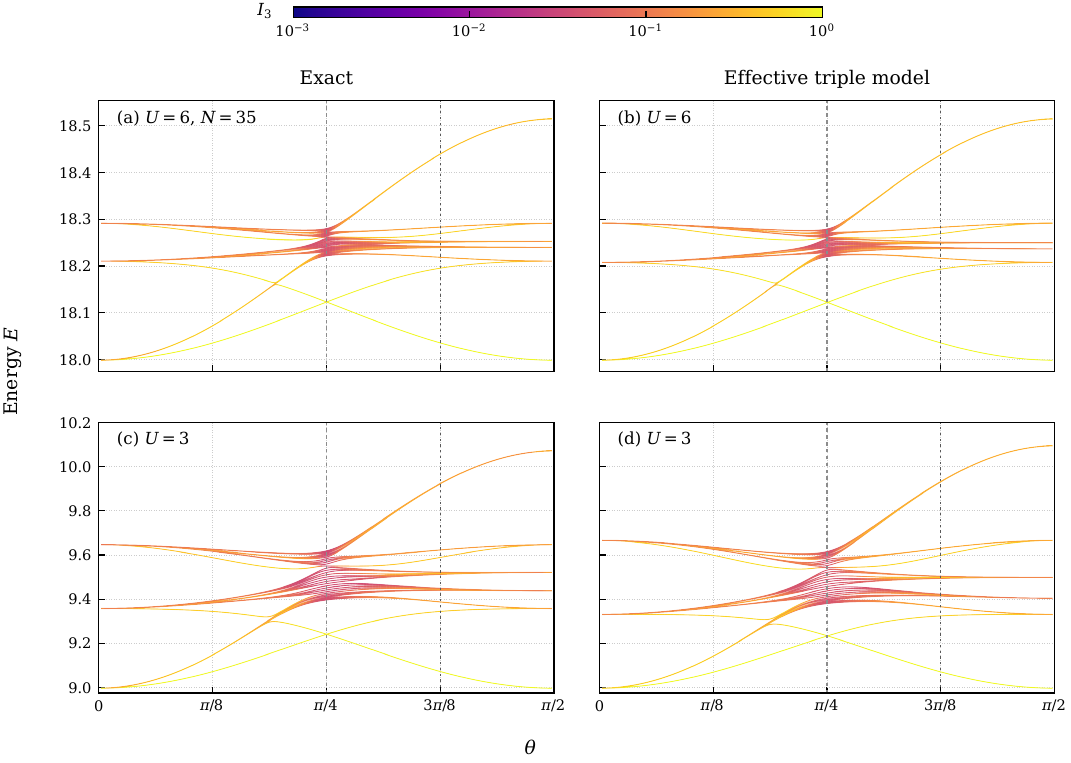}
\caption{Exact (left) and effective (right) triplon spectra for $S_8$, $N=35$, at $U=6$ (top) and $U=3$ (bottom). Sparse diagonalization () gives the highest 35 eigenstates of the full 7770-dimensional three-boson Hamiltonian; one additional eigenvalue checks separation from the remaining spectrum. In the basis $|T_j\rangle=(b_j^\dagger)^3|0\rangle/\sqrt6$, the effective Hamiltonian has diagonal energy $3U+3(t_{j-1}^2+t_j^2)/(2U)$ and hopping $3t_j^3/(2U^2)$, with missing end bonds set to zero. Colors encode $I_3=\sum_{\boldsymbol n}|\langle\boldsymbol n|\psi\rangle|^4$, summed over the full normalized three-boson Fock basis for the exact states. The effective eigenvectors are embedded in the same-site trion configurations, with other amplitudes zero. The common logarithmic scale is not rescaled separately in each panel. Energy limits are shared within each row. Each sweep contains 159 interior angles; color values below $10^{-3}$ are clipped. Colors near unresolved degeneracies can depend on the numerical eigenbasis.}
\label{fig:triple_comparison}
\end{figure}

\clearpage
\begin{figure}[htbp]
\centering
\includegraphics[width=\textwidth,height=.68\textheight,keepaspectratio]{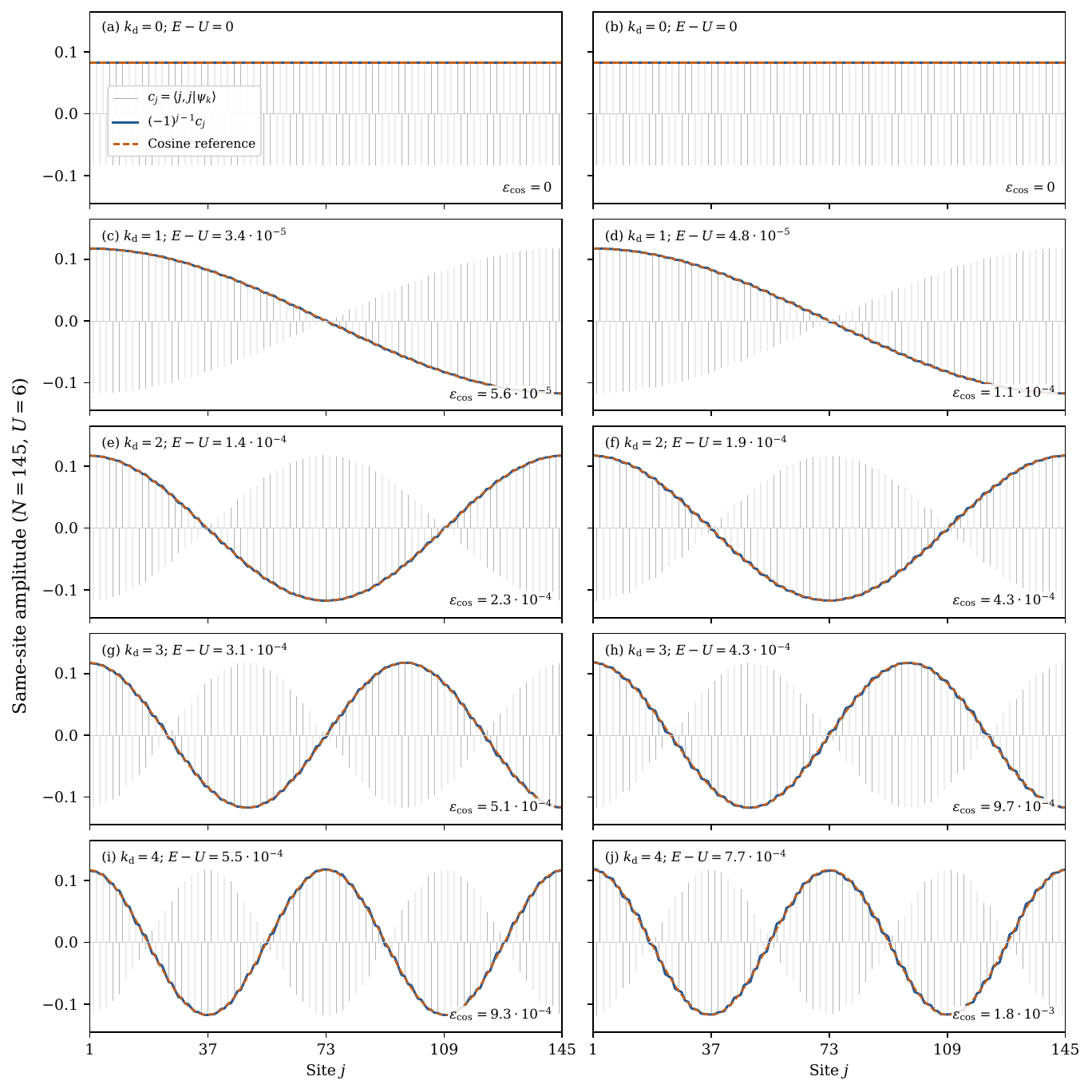}
\caption{Lowest doublon states for $N=145$ and $U=6$, at $\theta=\pi/8$ (left) and
$3\pi/8$ (right). Rows show the doublon-band indices $k_{\rm d}=0,\ldots,4$.
Gray stems show the signed same-site coefficients $\langle j,j |\psi_k\rangle$
of the exact two-boson eigenstates. Blue curves show their envelopes after removing
the alternating sign, $f_j=(-1)^{j-1}\psi_{jj}$. Dashed curves are the free-end cosine
profiles $h_m(j)\propto\cos[m\pi(j-1/2)/N]$, with $m=k_{\rm d}$, scaled to the
same norm as $f$. The annotated cosine-envelope error is
$\epsilon_{\cos}=1-|\sum_j h_m(j)f_j|^2/P_d$, where $h_m$ has unit norm and
$P_d=\sum_j|\psi_{jj}|^2$. The exact $E=U$ state has a constant envelope; the adjacent
higher-energy states exhibit approximately harmonic envelopes with small
quasiperiodic deviations. All numerical coefficients retain full-state normalization.}
\label{fig:low_doublon_profiles}
\end{figure}